\documentclass[]{zakoproc}
\usepackage{graphicx}
\begin{document}

\title{The interplay between cosmic rays and magnetic turbulence in galaxy 
clusters: radio halos and $\gamma$--rays}
\author{Gianfranco Brunetti}
\institute{INAF- Istituto di Radioastronomia, via P. Gobetti 101, I--40129
Bologna, Italy}
\markboth{G. Brunetti}{Radio emission and $\gamma$--rays from galaxy clusters}

\maketitle

\begin{abstract}
The interaction of magnetic turbulence and relativistic particles is a
important process for understanding particles propagation and acceleration
in many astrophysical environments.
Large-scale turbulence can be generated in the intra-cluster-medium (ICM)
during mergers between galaxy clusters and affects their 
non-thermal properties. 
Giant radio halos, Mpc-scale synchrotron sources observed in merging
clusters, may probe the connection between turbulence and non-thermal
cluster-scale emission.
After discussing relevant aspects of the physics of turbulence and
turbulent acceleration
in the ICM, I will focus on recent advances
in the modeling of non-thermal emission from galaxy clusters.
\end{abstract}

\section{Introduction}

Clusters of galaxies and filaments are the largest structures in the 
present universe in which the gravitational force due to dark matter overcomes
the expansion of the universe. 
Baryonic matter in clusters is heated to roughly the virial
temperature, but there is room to accommodate a non negligible amount 
of non-thermal energy in the form of accelerated particles. 
Indeed clusters are expected to be sites of acceleration and 
storage of charged particles and sources of non-thermal radiation from 
radio to $\gamma$--rays (Blasi et al 2007 for review).

This theoretical picture is supported by radio observations that 
indeed show the presence of diffuse synchrotron radiation on Mpc scales 
in a fraction of massive and nearby galaxy clusters.
The radio emission is due to relativistic electrons, with energies several
GeV, diffusing in $\mu$G magnetic fields and it is classified as  
{\it radio halos}, roundish low brightness sources from cluster central 
regions, and {\it radio relics}, elongated sources typically found in 
the clusters outskirts (e.g., Ferrari et al. 2008 for review).
Giant radio halos are the most spectacular and best studied cluster-scale 
non-thermal sources. Limits on their polarization (at few percent level) 
and their morphological similarity with the cluster X-ray emission 
suggest 
that the radio emission is generated from the same regions
where X-rays are produced, and that 
the relativistic plasma is mixed with the thermal one.
Radio observations and their follow up in the X-rays show that 
giant radio 
halos are always found in merging systems while they are not generated 
in more relaxed clusters, suggesting that a fraction of 
the gravitational energy that is dissipated during merger events 
is channelled into the generation of non-thermal components (eg. Cassano et
al. 2010 and ref therein). 

\noindent
Two important mechanisms may play a role for the origin of radio halos:
the generation of secondary electrons from inelastic collisions between
relativistic and thermal protons (pp) in the ICM (eg.,Blasi \& Colafrancesco 1999, 
Pfrommer \& En\ss lin 2004, Keshet \& Loeb 2010), and the 
(re)acceleration of primary and/or secondary particles in the
ICM due to turbulence generated during cluster mergers (eg. Brunetti et al 2001,
Petrosian 2001, Brunetti \& Blasi 2005).
Turbulent reacceleration may naturally explain the connection between
halos and cluster mergers and became a popular scenario for the origin of 
radio halos. Most important 
the synchrotron spectrum of a number of radio halos shows 
a curvature at higher frequencies (or it is very steep) suggesting that
the mechanisms responsible for the acceleration of 
electrons in the emitting $Mpc^3$-regions are
poorly efficient, with acceleration time
$\tau_{acc} \sim 10^8$yrs, consistent with acceleration by turbulence  
(Schlickeiser et al 1987, Brunetti et al 2008).

In this contribution I will focus on relevant aspects
of turbulence and particle acceleration by magnetic turbulence in the ICM and on the
clusters spectral energy distribution expected from models of turbulent
acceleration of primary and secondary particles.

\section{Turbulence in galaxy clusters}

A simple estimate of the Reynolds number in the ICM gives
${\cal Re}\sim 10^2$ which is barely sufficient to initiate
turbulence. This value however significantly increases if 
the effect of the ICM-magnetic field (mean and rms) 
on the particle mean free path (and viscosity) is 
taken into account, suggesting that the ICM is turbulent at some level
(eg., Subramanian et al 2006, Brunetti \& Lazarian 2007).
The properties of turbulence in galaxy clusters are discussed in many
review papers (eg Lazarian \& Brunetti 2011 for recent review), in the
following I will briefly describe few observational and theoretical aspects.

Limits to the turbulent velocity support in the cores of cool-core
clusters have been recently obtained through X-ray spectroscopic
observations, they constrain the energy of turbulent motions  
on 10-30 kpc scales at $\sim 10-20$\% of the thermal energy
budget (Sanders et al 2011).
Analysis of the pressure fluctuations in the cores of a few galaxy
clusters agree on the possibility of a substantial pressure
support from turbulent motions, of the order of 10\% of the thermal
pressure (eg., Churazov et al 2011 and ref therein).
Direct evidence of turbulent motions in the ICM come
from the analysis of the Faraday Rotation (RM) of the polarised emission from
cluster and background radio sources. The magnetic field topology inferred
by these studies is turbulent with coherent scales ranging from a few kpc
to several ten of kpc (Bonafede et al 2010 and ref therein).

\noindent
The generation of turbulent motions on large scales in galaxy clusters
is a unavoidable byproduct of the hierarchical process 
of cluster formation (eg Iapichino et al 2008, 
Ryu et al. 2008, Vazza et al 2011).
Numerical simulations suggest that turbulence 
is produced mainly by merger-induced shear flows and that in the case
of merging clusters it may contribute up 
to 10-30\% of the total thermal energy budget.
Large-scale turbulent motions, injected at scales 
$L_o \sim 100-500$ kpc during mergers, are important drivers 
of turbulence at smaller scales.
We expect typical velocities of the turbulent eddies at large scales 
$\delta V_{L_o} \sim 300-700$ km/s which makes turbulence sub--sonic,
but strongly super--Alfv\'enic ($\delta V_{L_o} >> V_A$, $V_A$ the 
Alfven velocity in the ICM). 
Turbulence at large scales is thus essentially hydrodynamic and made of 
compressive and incompressive eddies.
Under these conditions lines of the mean field in the ICM are tangled 
and stirred by motions resulting 
in a complex topology of the ICM-field, in 
line with observations of RM.
Viscosity in a turbulent and magnetised ICM is strongly suppressed due
to the effect of the bending of magnetic field lines and of the
perturbations of the magnetic field at small scales induced by plasma
instabilities.
The important consequence is that turbulence is expected to establish 
a inertial range down to small scales. 
In the inertial range the velocity of turbulent eddies decreases 
with scales and becomes sub-Alfvenic, at these scales the properties
of magnetic turbulence in the ICM should be similar to those of MHD
turbulence.

\noindent 
At small scales the ICM becomes collisionless and additional physics
comes into play. The non-linear coupling between turbulent and
particles drains a fraction of turbulent energy into particle
heating and acceleration.
In the classical formulation the ICM becomes collisionless at
scales $\leq$ the Coulomb scale $\sim 10$ kpc.
However the effective mean free path in a turbulent plasma may be determined
by other complex effects, for example by the scattering with magnetic field
perturbations generated by plasma instabilities.
Although this territory is still poorly explored,
we might reasonably claim that the ICM ``behaves'' collisional at 
scales significantly smaller than the classical Coulomb scale (Brunetti
\& Lazarian 2011a).
The generation of plasma instabilities due to the coupling between turbulent
modes and thermal and relativistic particles is expected to back react
on the turbulent spectrum generating waves at small
scales. This provides a elegant mechanism, complementary
to turbulent cascading, to transport turbulent
energy from large to very small scales (eg. Yan \& Lazarian 2011).

\section{Particle acceleration by magnetic turbulence}

Charged particles in magnetic turbulence are accelerated stochastically
by interacting with electric and magnetic field fluctuations (eg. Melrose 1980 for review).
The condition for resonant interaction between a particle with
momentum $p$ and a wave with frequency $\omega$ and wavenumber $k$ is
$\omega = k_{\vert\vert} v_{\vert\vert} + n \Omega/\gamma$ ($k_{\vert\vert}$ and
$v_{\vert\vert}$ are the wavenumber and particle-velocity components along
the mean magnetic field), where $n = \pm 1, 2, ..$ give gyroresonance
with electric field fluctuation. The case $n=0$ marks Transit-Time-Damping (TTD),
a coupling between the particle magnetic-momentum and
the magnetic (turbulent) parallel gradients.

\noindent
Acceleration of electrons from the thermal pool to relativistic energies
by MHD turbulence in the ICM faces serious problems due to energy arguments
(eg. Petrosian \& East 2008).
Consequently, turbulent acceleration models must assume
a pre-existing population of relativistic particles that
provides the seeds to reaccelerate during cluster mergers (eg.
Brunetti 2011 for review).
A correct description of the process of reacceleration of seed particles 
is challenging.
It requires taking into account the fundamental properties of 
magnetic turbulence as well as the mutual feedback of magnetic fields 
and energetic particles in the turbulent medium.
The last decade has been marked by substantial advances in understanding 
of magnetic turbulence in the MHD regime that provides a solid
``guide'' to model turbulence in galaxy clusters across
a fairly large range of spatial scales.
At collisionless scales the collisionless interaction of turbulent perturbations 
with thermal and non-thermal particles changes the turbulent spectrum 
with time since dampings transfer 
turbulent energy into particles, this in turn affects 
(self regulates) also the efficiency of the particle acceleration process.
The picture becomes more challenging when additional 
processes, such as plasma instabilities, are considered.
A variety of instabilities (e.g. firehose, mirror,
gyroresonance etc) can be generated in the ICM in the presence of
turbulence, they can lead to the transfer of energy from 
large-scale turbulence to perturbations on smaller scales and may 
play a role in the process of scattering and acceleration of relativistic
particles. A satisfactory modeling of the particle acceleration 
process by magnetic turbulence in galaxy clusters would require taking
into account self-consistently all these mechanisms, which is a 
challenging territory for future studies. Yet substantial steps in the
modeling of turbulent acceleration in galaxy clusters lead to conclude
that it is a important process to explain the phenomenology of non-thermal
emission from cluster-scale.

In Brunetti \& Lazarian (2007) we considered the advances in the
theory of MHD turbulence to develop a comprehensive picture of
turbulence in the ICM and to study the reacceleration of relativistic
particles by compressive turbulence by considering all the
relevant collisionless damping processes.
According to our approach large scale turbulence, generated in the
ICM during cluster mergers, cascades at smaller (collisionless)
scales where turbulent waves
can interact non-linearly with relativistic and thermal particles.
Under the hypothesis that the collisionless scale of the ICM is 
$\sim$Coulomb ion mean free path, particle acceleration is mainly
due to fast modes. The most important
collisionless damping of these modes in the hot ICM is the TTD resonance with thermal particles that
essentially transfers turbulent energy into heating of the thermal ICM.
In this case it is calculated that
only $\sim$ 10\% of the energy of compressible turbulence goes into the
(re)acceleration of seed relativistic particles via TTD.
This scenario allows prompt calculations of particle acceleration by
MHD turbulence in the ICM. 
The ensuing cluster-scale radio emission generated
in merging clusters results in very good agreement with present observations of radio halos.

\noindent
In other physical situations a larger fraction of the turbulent energy 
can be dissipated into the (re)acceleration of seed relativistic particles 
in galaxy clusters.
This is the case of the gyro-resonant interaction
with Alfv\'en modes at small scales that has been considered in several
papers that attempt modeling radio halos 
(e.g. Ohno et al 2002, Fujita et al. 2003, Brunetti et al. 2004).
A large fraction of the energy of compressible turbulence can also be
transferred to relativistic particles by TTD with 
fast modes
under the assumption that the ICM "behaves" collisional at scales
much
smaller than the Coulomb ion mean free path (e.g. Brunetti \& Lazarian
2011a).
In all these cases where a large fraction of turbulent energy is
channeled in the reacceleration of relativistic particles 
the efficiency of the particle acceleration process is self-regulated
by the back--reaction (damping) of particles on the spectrum 
of turbulence.
Stronger turbulence induces more efficient acceleration 
leading to a faster growth of the particles energy density with time.
This -- however -- increases the damping of turbulence 
and the interaction approaches a quasi--asymptotic (and very complex)
regime where relativistic particles get in (quasi) equipartition with 
turbulence and self-regulate their (re)acceleration.

\section{Turbulent (re)acceleration of primary and secondary particles}

Cosmic ray protons in the ICM are long--living particles that
can be confined (and accumulated) in
clusters (V\"olk et al 1996, Berezinsky et al 1997) with 
the unavoidable generation of secondary particles in the ICM.
Secondary electrons provide a natural reservoir of seed particles
that can be reaccelerated by interacting with magnetic turbulence
in galaxy clusters.
In Brunetti \& Lazarian (2011b) we model self-consistently the
interaction of compressible turbulence and relativistic 
protons and their secondaries generated in the ICM.

This general scenario predicts that galaxy clusters are non-thermal
sources from radio to $\gamma$--rays. The gamma-ray emission is mainly due to
decay of $\pi^o$ generated from pp collisions in the ICM and it does not
(strongly) depend on cluster dynamics.
The level of radio emission from these models is tightly connected with
the dynamical properties of the hosting clusters. Magnetic turbulence
in merging clusters reaccelerates secondary electrons generated from pp
collisions resulting in enhanced "on state" synchrotron emission in the
form of giant radio halos. The scenario leads to the unavoidable expectation 
of "off-state" radio halos in relaxed systems, namely Mpc-scale
emission
generated by the continuous injection of secondary electrons in the ICM. 
Calculations predict a luminosity of "off-state" halos in more relaxed systems 
roughly one order of magnitude smaller than that of "on state" radio halos,
a possibility that can be tested with deep radio surveys (Brown et al 2011).

Figure 1 shows the expected spectrum in the case of the Coma 
cluster where the energy content of compressible 
turbulence is assumed $\approx 18\%$ of the ICM 
and the energy density of the reaccelerated relativistic protons is 
$\sim 4 \%$ of the ICM. The model provides a very good agreement
with both the synchrotron spectrum and radio brightness distribution
of the radio halo (see caption for details). Gamma ray emission
is expected from the Coma cluster at $\sim 10-20 \%$ level of present 
upper limits from FERMI.
In Fig.1 we also show model results where ``no'' turbulence is
assumed. 
Under this condition radio emission is maintained only by the process
of continuous injection of secondary electrons from pp collisions.
In this case however to explain the radio luminosity and brightness profile
of the Coma halo it is necessary to assume a large energy content of 
primary relativistic protons, about 10 times larger than in the
previous case.
The drawback of this scenario is a $\gamma$--ray emission from $\pi^o$ decay
that is larger than present limit.
This suggests that the mechanism of generation of secondary particles
via pp collisions in the ICM (secondary models), when considered alone,
is not sufficient to explain radio halos, unless the cluster magnetic
field is substantially larger than that derived from RM (Brunetti et
al in prep).

\begin{figure*}
 \includegraphics[width=.32\textwidth]{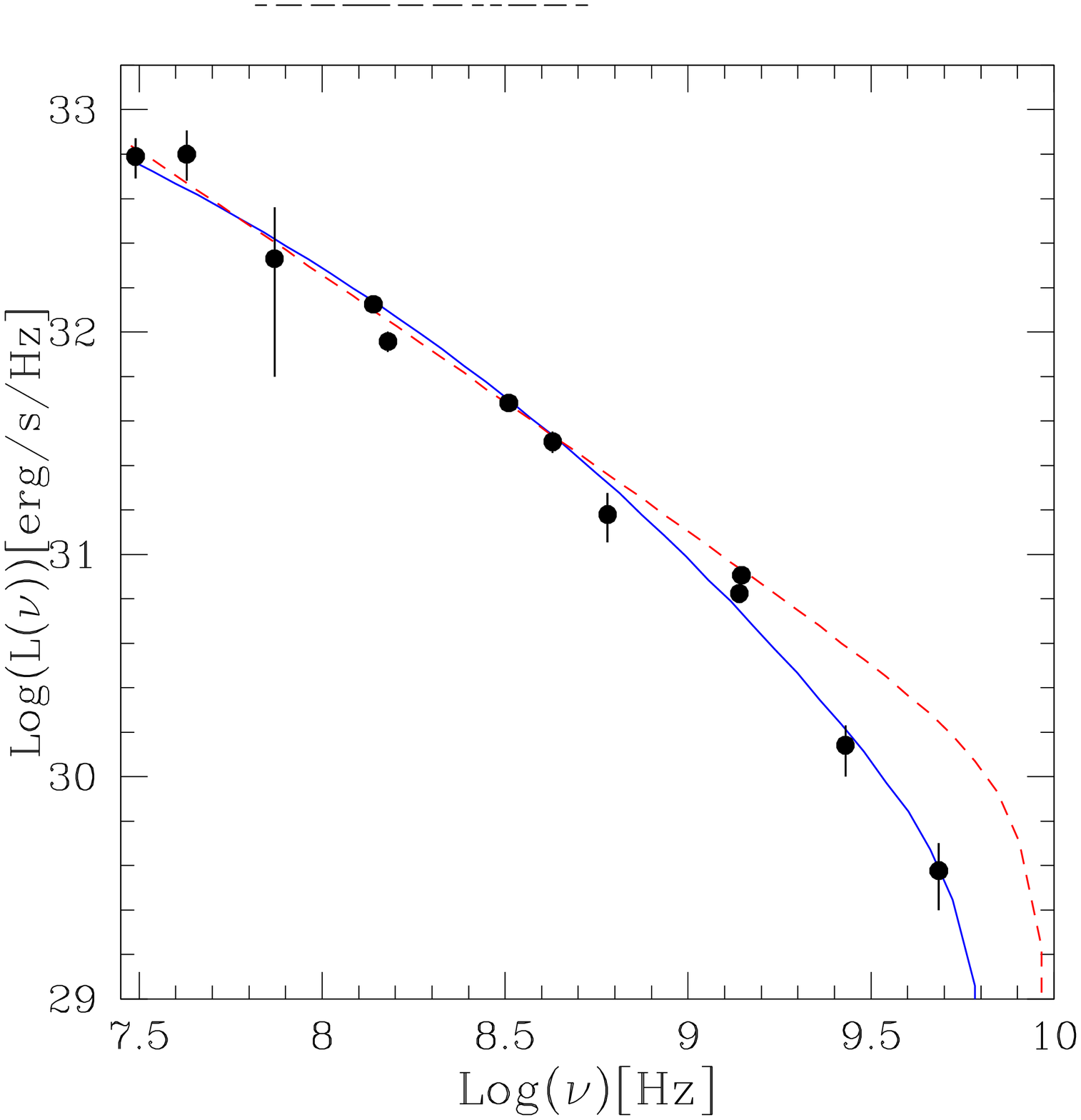}
 \includegraphics[width=.32\textwidth]{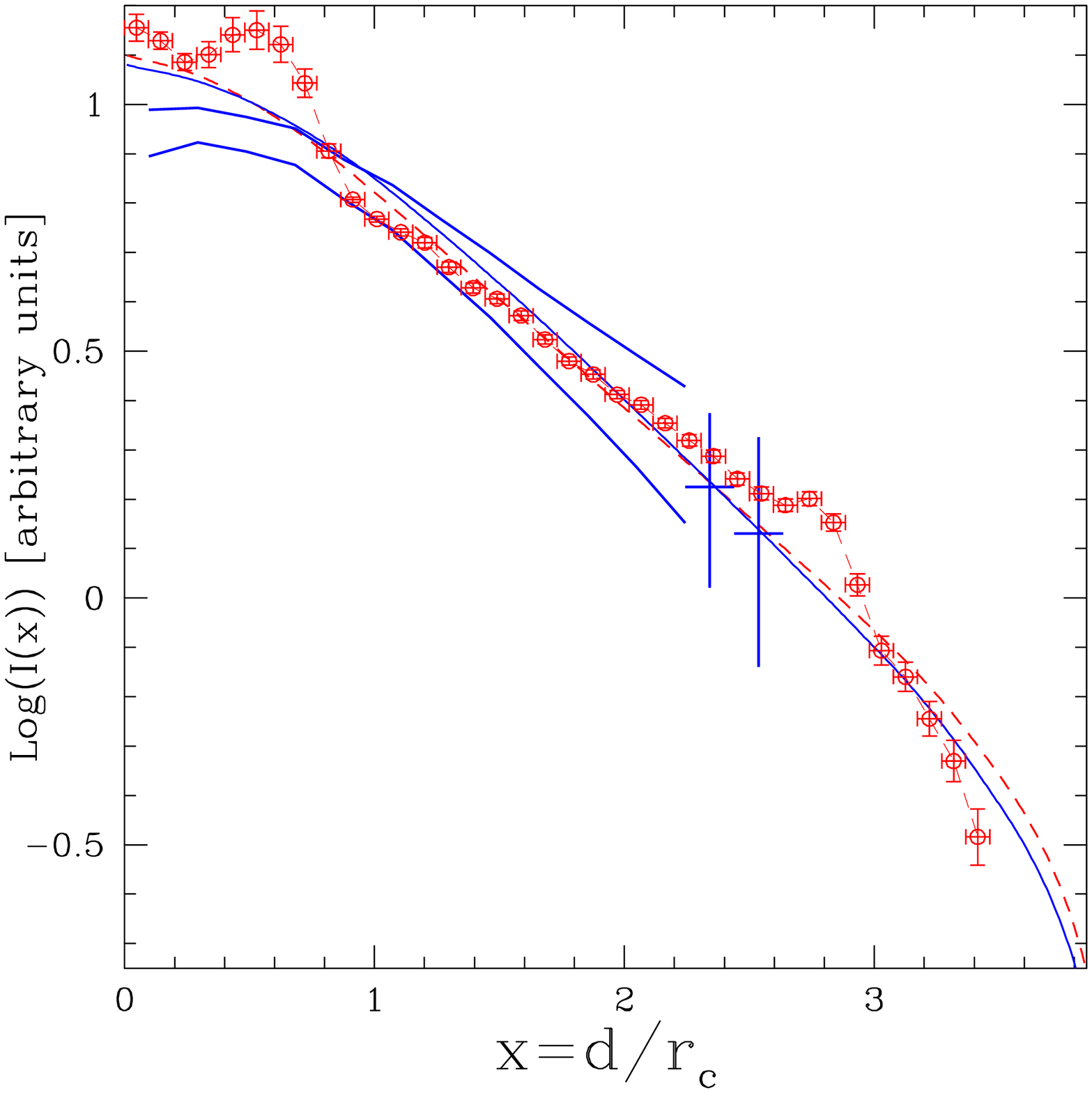}
 \includegraphics[width=.32\textwidth]{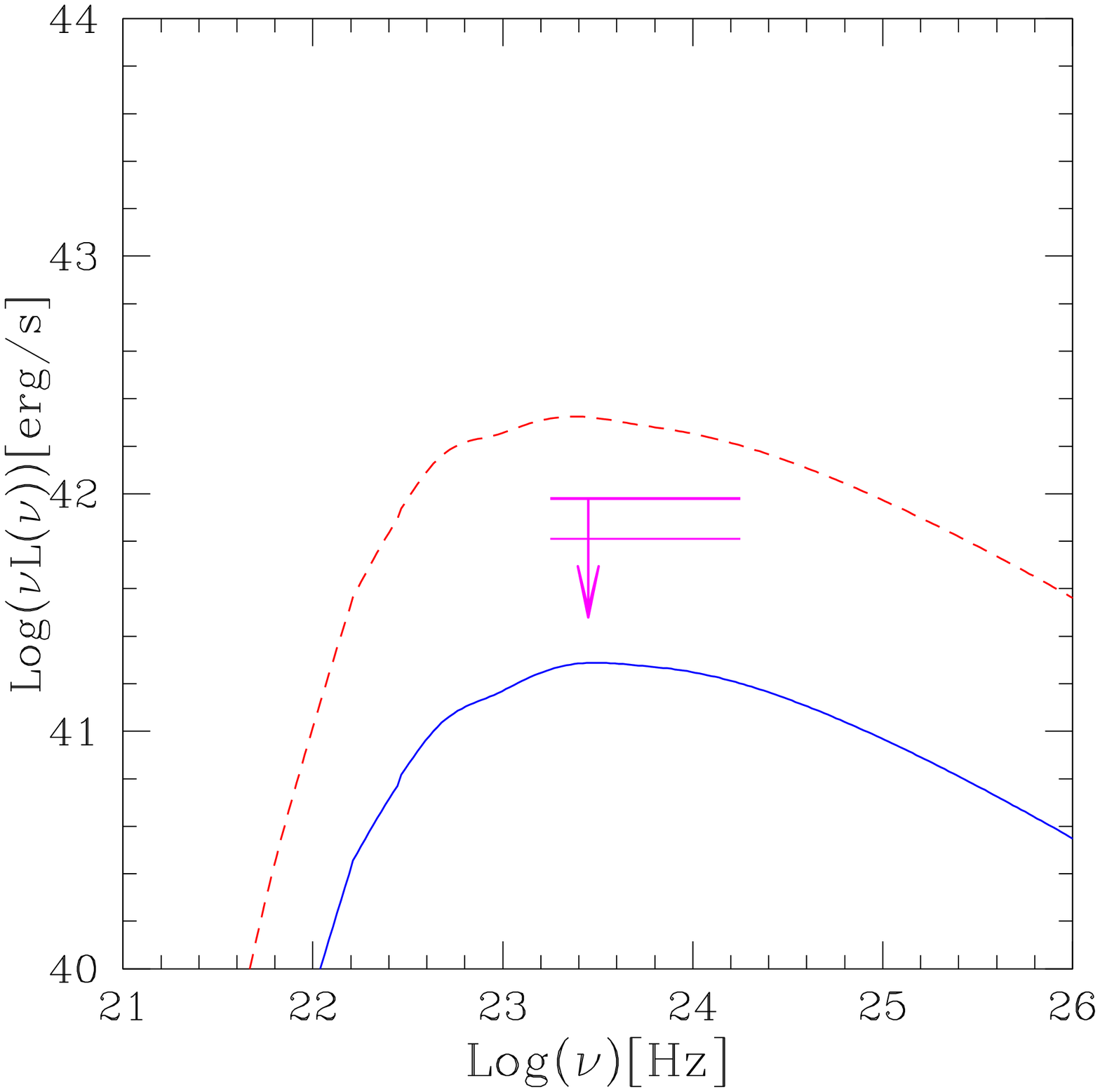}
 \caption{
 Synchrotron spectrum (left) and brightness profile (central panel; the radial distance is in units of 
 core radius, red points and blue data are from 325 MHz observations) of the Coma radio halo,
 and the cluster $\gamma$--ray spectrum from $\pi^o$-decay (right; the $2-\sigma$
 FERMI upper limit is from Ackerman et al 2010, we also show the FERMI sensitivity
 expected after 3.5 years).
 We show the case of turbulent reacceleration (solid-blue) and pure-secondary  (dashed-red)
 models. We assume a scaling between relativistic and thermal protons energy-densities
 $\epsilon_{CR} \propto \epsilon_{TH}^{1-f}$ with $f=1.15$ and 1.5 for reacceleration and
 secondary model, respectively.
 Following best fit from RM (Bonafede et al 2010) the magnetic field energy density is scaled with 
 the thermal  energy density and the central field value is $= 4.7 \mu$G.
 }
\end{figure*}

\section{Conclusions}

Giant radio halos are probes of complex mechanisms of particle acceleration
in turbulent ICM-regions. 
Substantial advances in understanding stochastic
particle acceleration by magnetic turbulence in galaxy clusters have been
achieved by using physically motivated models of turbulence and by considering 
the reacceleration of both primary and secondary particles in the ICM.
These studies suggest that turbulent acceleration is a important process
to understand
the phenomenology of radio halos and their connection with cluster mergers.
The lack of detections of galaxy clusters in the $\gamma$--rays provides additional
constraints to the nature of radio halos.
Turbulent reacceleration models are consistent with the radio properties of the Coma halo and with the 
present FERMI upper limit. On the other hand the brightness profile of the
radio halo would imply a $\gamma$-ray luminosity that is appreciably larger than the
FERMI limit assuming pure secondary models; in order to reconcile
secondary models with radio and $\gamma$-ray observations the magnetic energy
density in the Coma cluster must be postulated 5-10 times larger than
that constrained from RM.

\acknowledgements{The author acknowledges partial support from PRIN-INAF2009}


\begin{thebibliography}{}

\bibitem[..]{Ackermann2010} 
Ackermann M., et al. 2010, ApJ, 717, L71

\bibitem{Berezinsky1997}
Berezinsky V.S., Blasi P., Ptuskin V.S., 1997, ApJ, 487, 529

\bibitem{Blasi1999}
Blasi P., Colafrancesco S., 1999, APh, 12, 169

\bibitem{Blasi2007}
Blasi P., Gabici S., Brunetti G., 2007, IJMPA 22, 681

\bibitem{Bonafede2010}
Bonafede A., et al., 2010, A\&A, 513, 30

\bibitem[]{} 
Brown, S., et al., 2011, ApJL, 740, 28

\bibitem[]{}
Brunetti, G, 2011, MmSAI 82, 515

\bibitem{Brunetti2001}
Brunetti, G., et al., 2001, MNRAS, 320, 365

\bibitem{} Brunetti G., et al., 2004, MNRAS 350, 1174

\bibitem[]{}
Brunetti, G., Blasi, P., 2005, MNRAS, 363, 1173

\bibitem[Brunetti \& Lazarian (2007)]{BL2007} 
Brunetti G., Lazarian A., 2007, MNRAS, 378, 245

\bibitem[Brunetti et al. (2008)]{Brunetti2008} 
Brunetti, G., et al.\ 2008, Nature, 455, 944

\bibitem[{Brunetti \& Lazarian (2011a)}]{BL2011a}
Brunetti G., Lazarian A., 2011a, MNRAS, 412, 817

\bibitem[{Brunetti \& Lazarian (2011b)}]{BL2011b}
Brunetti G., Lazarian A., 2011b, MNRAS, 410, 127

\bibitem[..]{..} Cassano, R., et al., 2010, ApJL, 721, L82

\bibitem[]{} Churazov, E., et al., 2011, arXiv1110.5875

\bibitem{} Ferrari, C. et al. 2008, SSRv 134, 93

\bibitem{Fujita2003} Fujita Y., Takizawa M., Sarazin C.L., 
2003, ApJ, 584, 190\bibitem{Keshet2010} Keshet R., Loeb A., 2010, ApJ, 722, 737

\bibitem[Iapichino \& Niemeyer (2008)]{Iapichino2008} 
Iapichino, L., \& Niemeyer, J.~C.\ 2008, MNRAS, 388, 1089 

\bibitem[]{} 
Lazarian A., Brunetti, G., 2011, MmSAI, 82, 636

\bibitem[Melrose(1980)]{1980panp.book.....M} Melrose, D.~B., 1980, {\it Plasma astrohysics. Nonthermal processes in diffuse magnetized plasmas},
New York: Gordon and Breach

\bibitem{} Ohno, H., Takizawa, M., Shibata, S., 2002, ApJ, 577, 658 

\bibitem{Petrosian2001} Petrosian V., 2001, ApJ, 557, 560 

\bibitem{Petrosian2008} Petrosian V., East W.E., 2008, ApJ, 682, 175

\bibitem[]{} Pfrommer, C., Ensslin, T.A., 2004, MNRAS, 352, 76

\bibitem[]{} Ryu, D., et al., 2008, Science,  320, 909

\bibitem[Sanders et al. (2011)]{Sanders2011} 
Sanders, J.~S., Fabian, 
A.~C., \& Smith, R.~K.\ 2011, MNRAS, 410, 1797 

\bibitem{Schlickeiser1987} Schlickeiser R., et al., 1987, A\&A, 182, 21

\bibitem[]{}
Subramanian, K., et al., 2006, MNRAS, 366, 1437

\bibitem[Vazza et al. (2011)]{Vazza2011} 
Vazza, F., et al., 2011, A\&A, 529, A17 

\bibitem{Voelk1996} V\"{o}lk H.J., et al., 1996, SSRv, 75, 279

\bibitem[Yan \& Lazarian (2011)]{YanL2011} 
Yan, H., \& Lazarian, A.\ 2011, ApJ, 731, 35

\end{thebibliography}
\end{document}